\documentclass[5p,times]{elsarticle}

\usepackage{graphicx}  % Include figure files
\usepackage{subfigure}
\usepackage{multirow}
\usepackage{float}
\usepackage{tikz}
\usepackage{xfrac}
\usepackage{comment}
\usepackage{dirtytalk}
\include{tikzspecs}
\include{booktabs}
\usepackage{blindtext}
\usepackage{cuted}
\usepackage{float}

\usepackage{hyperref}
\hypersetup{
colorlinks=true,
urlcolor= blue,
citecolor=blue,
linkcolor= blue,
}

\usepackage{fancyhdr}
\usepackage{longtable}
\usepackage{parskip}
\usepackage[T1]{fontenc}
\usepackage{dcolumn}   % Align table columns on decimal point

\usepackage{bm}        % bold math
\usepackage{amsfonts}  % Common math fonts
\usepackage{amsmath}   % Common math functions
\usepackage{amssymb}   % Common math symbols
\usepackage{booktabs}
\usepackage{cleveref}
\usepackage{balance}
\crefformat{figure}{#2{\color{black}#1}#3}
\crefformat{table}{#2{\color{black}#1}#3}

\usepackage{xcolor}
\definecolor{gabby}{rgb}{0,0.0,0.4}

\newcommand{\pwiseout}{\end{array}\right.}

\newcommand{\Plus}[0]{\texttt{+}}

\setcitestyle{number,open = {[},close={]}}

\begin{document}
\title{System Size and Flavour Dependence of Chemical Freeze-out Temperatures in ALICE Data from pp, pPb and PbPb Collisions at LHC Energies}
\author[1,2]{Fernando Antonio Flor}
\ead{fernando.flor@yale.edu}
\author[2]{Gabrielle Olinger}
\author[2]{Ren\'e Bellwied}
\address[1]{Wright Laboratory, Yale University, New Haven, Connecticut 06520, USA}
\address[2]{Department of Physics, University of Houston, Houston, Texas 77204, USA}
\date{\today}

\begin{abstract}
We present the system size and flavour dependence of the chemical freeze-out temperature ($T_\mathrm{ch}$) at vanishing baryo-chemical potential calculated via thermal fits to experimental yields for several multiplicity classes in pp, pPb and PbPb collisions measured by ALICE. Using the Thermal-FIST Hadron Resonance Gas model package, we compare the quality of fits across various treatments of strangeness conservation under different freeze-out conditions as a function of the charged particle multiplicity density $\big \langle dN_\mathrm{ch}/d\eta \big \rangle$. Additionally, we examine how the anti-hadron to pion yield ratios of light and strange baryons, as well as the $\phi$ meson, evolve within a flavour-dependent model. Through a unique two-temperature chemical freeze-out approach, we show that flavour dependence of $T_\mathrm{ch}$ in a Strangeness Canonical Ensemble leads to a natural explanation of strangeness enhancement from small to large systems at LHC energies without requiring any non-equilibrium particle production at small $\big \langle dN_\mathrm{ch}/d\eta \big \rangle$.
\end{abstract}


\maketitle
\setlength{\parskip}{0em}

\section{Introduction}

\indent Hadronization and chemical freeze-out have been suggested to coincide at the phase boundary in the Quantum Chromodynamics (QCD) phase diagram based on results from Statistical Hadronization Models (SHMs) using particle yields measured at the Large Hadron Collider (LHC) and the Relativistic Heavy Ion Collider (RHIC), when compared to pseudo-critical temperature calculations from temperature dependent continuum extrapolations of the chiral susceptibilities on the lattice \cite{Stachel_2014, HotQCD1, WB2, WB1, Lattice}. A point of interest emerges concerning whether the phase transition from quark to hadron degrees of freedom occurs at the same temperature for all particle species
and/or quark flavours.\\
\indent Final state particle yields have been successfully reproduced by SHMs to nine orders of magnitude over a wide energy range in high energy collisions of heavy ions \cite{PBMNature}. SHMs typically use experimental hadron yields from central events (0 - 10\%) in relativistic heavy-ion collisions as an anchor for determining common freeze-out parameters in the QCD phase diagram -- namely, the chemical freeze-out temperature ($T_\mathrm{ch}$) and the baryo-chemical potential ($\mu_\mathrm{B}$) -- within a Grand Canonical Ensemble (GCE), where baryon number, electric charge and strangeness are conserved on average.\\  
\indent The GCE treatment has been shown to inadequately reproduce experimental results in a multiplicity-dependent manner across pp, pPb and PbPb collision systems measured by the ALICE Collaboration -- particularly where  $\big \langle dN_\mathrm{ch}/d\eta \big \rangle \leq 20$ \cite{SharmaPRC1}. The aforementioned fact is often attributed to the presence of non-equilibrium strangeness production in the smaller systems and may be partially remedied by employing a Strangeness Canonical Ensemble (SCE), in which strangeness is explicitly conserved within a correlation volume $V_\mathrm{C}$ using a single value of $T_\mathrm{ch}$ for all particle species (1CFO). The question arises as to whether out-of-equilibrium strangeness in small systems provides an adequate description of the overall production of final state (multi)- strange hadrons as measured at LHC energies. The presence of said disequilibrium, commonly represented by a  non-negligible strangeness saturation parameter ($\gamma_\mathrm{{S}}$) within the fireball volume of hadronic collisions, is known to diminish with increasing collision energy \cite{Tounsi_Canonical_Suppression1, Tounsi_Canonical_Suppression2}. 

In principle, strangeness equilibration scaling as an inverse function of collision energy can be equated with an increasing value of $\gamma_\mathrm{{S}}$. At LHC energies, $\gamma_\mathrm{{S}}$ is expected to be asymptotic to unity \cite{SharmaPRC1, Biswas_HRG1}, i.e. full saturation of strangeness is achieved as a function of $\big \langle dN_\mathrm{ch}/d\eta \big \rangle$. The dominating presence of strangeness enhancement at ALICE, even in small systems, is evident from an energy dependent comparison of final state anti-hadron to $\pi^{+}$ yields measured by ALICE and STAR \cite{RB_SequentialFO}, where production of strange baryons is fully saturated for collision energy values above $\sqrt{s_{\rm{NN}}} = 62.4$ GeV.

Nevertheless, in order to fully describe the experimental data, further considerations are made regarding the interplay between $V_\mathrm{C}$, the fireball volume ($V$), the experimental rapidity window ($\Delta$y) and $\gamma_\mathrm{{S}}$ \cite{VV_Donigus_Stoeker_PRC1, ALICE_pPb_Fits}. This letter aims at providing a description of strangeness enhancement across all three collision systems measured by ALICE, assuming full equilibration and saturation of strangeness is inherently present at LHC energies. Our description of final state hadron yields within the SHM framework relies solely on flavour-dependent freeze-out temperatures and volumes across increasing $\big \langle dN_\mathrm{ch}/d\eta \big \rangle$ values. 

\section{Sequential Strangeness Freeze-out}
Flavour-dependent freeze-out temperatures in the crossover region of the QCD phase diagram have been predicted by continuum extrapolated susceptibilities of single flavour quantum numbers on the lattice \cite{Ratti2012, SuscRat}. Flavour specific susceptibility ratios ${\chi_{4}}/{\chi_{2}}$, suggested as an observable for directly determining freeze-out temperatures \cite{Karsch2012}, show a deviation of the lattice and Hadron Resonance Gas (HRG) model calculations at the peaks of the lattice data, which occur at flavour-dependent temperatures differing by $15 - 20$ MeV from light to strange quarks \cite{SuscRat}. Similar temperature differences between the light and strange mesons have also been shown in net-particle fluctuation measurements by the STAR Collaboration \cite{Adamczyk:2013dal, NetK, BluhmKaon, STAR:2020ddh}. An HRG-based study was also performed in a similar analysis on both off-diagonal and diagonal second order correlators of conserved charges \cite{Bellwied_PRD_2020}.

\indent In a previous letter \cite{Flor_etal_PLB20}, at vanishing $\mu_{\rm{B}}$, we calculated a light flavour freeze-out temperature $T_\mathrm{L} = 150.2 \pm 2.6$ MeV and a strange flavour freeze-out temperature $T_\mathrm{S} = 165.1 \pm 2.7$ MeV, by employing the GCE approach to heavy-ion collisions within the framework of the Thermal-FIST \cite{FIST} HRG model package and varying the particle species included in each thermal fit. 

\indent In this letter, we extend this flavour-dependent two chemical freeze-out (2CFO) temperature approach to pp, pPb and PbPb collision systems measured by ALICE as a function of increasing  $\big \langle dN_\mathrm{ch}/d\eta \big \rangle$, rendering a natural explanation of strangeness enhancement from small to large systems at vanishing $\mu_{\rm{B}}$.

\section{Model and Data Preparation}
All calculations shown here are performed using the open source Thermal-FIST thermal model package. In this iteration, we model an ideal non-interacting gas of hadrons and resonances within both, the GCE and the SCE scenarios, for the sake of comparison. The analysis is two-fold: to gauge the sensitivity of the chemical freeze-out temperature at vanishing $\mu_{\rm{B}}$ relative to the ensemble of choice, and to employ a 2CFO treatment onto the reigning ensemble. 

As the HRG input list, we use the PDG2016\Plus{} hadronic spectrum \cite{PDG2016+}. The PDG2016\Plus{} hadronic spectrum has been shown to be an optimized compromise between too few and too many resonant states when compared to lattice QCD predictions \cite{PDG2016+}; it includes a total of 738 states (i.e. *, **, *** and **** states from the 2016 Particle Data Group Data Book \cite{PDG16}).

Yield data for $\pi^{\text{+}}$, $ \pi^{\text{-}}$, $K^{\text{+}}$, $K^{\text{-}}$, $p$,  $\bar{p}$, $\Lambda$, $\bar{\Lambda}$, $\Xi^{-}$, $\bar{\Xi}^{+}$, $\Omega^{-}$, $\bar{\Omega}^{+}$, $K_{\rm{S}}^{0},$ and $\phi$ for ALICE pp collisions at $\sqrt{s} = 7.00$ TeV \cite{ALICENaturePhys}, pPb collisions at $\sqrt{s_{\rm{NN}}} = 5.02$ TeV \cite{ALICEpPb502_piKp, ALICEpPb502_phi, ALICEpPb502_MultiS} and PbPb collisions at $\sqrt{s_{\rm{NN}}} = 2.76$ TeV \cite{PbPb276,PbPb276K0S,PbPb276phi,PbPb276MultiS} across all available multiplicity classes are included in our analysis -- our multiplicity binning is shown in Table \cref{tab:binning}. Throughout this entire analysis, the yields for each particle and its corresponding anti-particle are assumed to be identical -- this methodology is explicitly employed in the case where the available experimental data only presents the sum of both particle and anti-particle yields. 

\begin{table}[htbp]
\centering
\caption{Available event centralities and corresponding values of charged particle multiplicity density for ALICE pp collisions at $\sqrt{s} = 7.00$ TeV, pPb collisions at $\sqrt{s_{\rm{NN}}} = 5.02$ TeV and PbPb collisions at $\sqrt{s_{\rm{NN}}} = 2.76$ TeV. For the pp sample, the multiplicity classes are labelled in accordance to their generalized definitions in \cite{ALICENaturePhys}.}

\label{tab:binning}
\begin{tabular}{@{}lll@{}}
\toprule

\multicolumn{3}{l}{\textbf{pp at 7.00 TeV}}    \\
Multiplicity Class & Event Centrality %\sigma/\sigma_{\sigma_\mathrm{INEL}>0} 
& $\langle dN_\mathrm{ch}/d\eta\rangle$ \\ \midrule
I-II               &      & 17.47$\pm$ 0.524                      \\
III-VI             &            & 10.383$\pm$0.313                      \\
VII-VIII           &          & 6.057$\pm$0.19                        \\
IX-X               &         & 2.886$\pm$0.135                       \\ \midrule
\multicolumn{3}{l}{\textbf{pPb at 5.02 TeV}}\\
%Centrality         & $\langle dN_\mathrm{ch}/d\eta\rangle$  &             \\
\midrule
&              & 45$\pm$1                       \\
&             & 36.2$\pm$0.8                   \\
&            & 30.5$\pm$0.7                   \\
&           & 23.2$\pm$0.5                  \\
&            & 16.1$\pm$0.4                   \\
&            & 9.8$\pm$0.2                    \\
&           & 4.3$\pm$0.1                    \\ \midrule
\multicolumn{3}{l}{\textbf{PbPb at 2.76 TeV}}\\
%Centrality         & $\langle dN_\mathrm{ch}/d\eta\rangle$  &             \\
\midrule
& 0-10$\%$            & 1447.5$\pm$54.5              \\
& 10-20$\%$           & 966$\pm$37                     \\
& 20-40$\%$           & 537.5$\pm$19                   \\
& 40-60$\%$           & 205$\pm$7.5                    \\
& 60-80$\%$           & 55.5$\pm$3                         \\ \bottomrule
\end{tabular}
\end{table}

All throughout, we follow a shorthand notation when naming our fits to (anti-)particle species (e.g. $\Omega$ refers to both $\Omega^{-} $ and $\bar{\Omega}^+$, etc.), unless explicitly noted otherwise.

In the GCE configuration, the thermal fits are performed with $T_{\rm{ch}}$ (MeV) and V (fm$^{3}$) as free parameters, fixing $\mu_{\rm{B}} = 0$, and setting $\gamma_{\mathrm{S}}$ and $\gamma_{\mathrm{q}}$ to unity in order to ensure a full saturation of strangeness and electric charge. Systematic HRG-based studies on the determination of the latter three parameters at top LHC energies can be found in Refs. \cite{Biswas_HRG2, Biswas_HRG3}.  We focus on varying the particle species included in the fit in order to gauge the sensitivity of $T_{\rm{ch}}$ to each fit. The particle species included in our flavour-dependent temperature fits are $\pi K p$, $\pi K p \Lambda \Xi \Omega K^{0}_{\rm{S}} \phi$ and $K \Lambda \Xi \Omega K^{0}_{\rm{S}} \phi$, hereinafter referred to as \say{light}, \say{all} and \say{strange}, respectively. We compare the extracted $T_{\rm{ch}}$ values, for the light, strange and all fits, and their corresponding $\chi^{2}$/dof measures as a function of their $ \big \langle dN_\mathrm{ch}/d\eta \big \rangle $ values across all three collision systems. Since the kaon yields have been shown to be insensitive to the freeze-out temperature \cite{Magestro_2002}, we include them in light fit for the sake of having sufficient degrees of freedom. As a proof of concept cross-check, we perform an independent fit replacing the kaon yields with (anti-)deuteron yields in the light fit for the most central (0 - 10\%) PbPb collisions at $\sqrt{s_{\rm{NN}}} = 2.76$ TeV \cite{ALICEDeuterons} and find no significant  difference in the extracted freeze-out parameters and the quality of fit -- shown in Table \cref{tab:thermfitdeu}.  Nevertheless, the present study does not implement the aforementioned $\pi p d$ light fit in our calculations since experimental yields for (anti-)deuteron yields are not yet available for all multiplicity bins across all three collision systems at ALICE. 

In the SCE configuration, our thermal fits are instead performed with $T_{\rm{ch}}$ (MeV) and V (fm$^{3}$) as free parameters, keeping $\mu_{\rm{B}} = 0$, $V_\mathrm{C}=V$ and setting $\gamma_{\mathrm{S}}$ and $\gamma_{\mathrm{q}}$ to unity. Setting $V_\mathrm{C}=V$ is done in order to guarantee a local conservation of strangeness within the calculated fireball volume per unit rapidity. It is worth mentioning that in this methodology $V_\mathrm{C}$ merely employs a strict conservation of the the strangeness quantum number and not of baryon number, which is treated grand canonically. This employment of global conservation of baryon number by our model in the SCE approach can be justified by the net-proton fluctuation data from the ALICE Collaboration \cite{ALICENetProtonFluctuations} -- assuming uniformity across all multiplicity classes at LHC energies -- indicating global baryon number conservation over a rapidity range significantly larger than unity.

The initial SCE analysis follows the same procedure than the early GCE trial, however, the following procedures are not performed for the GCE approach due to the deterioration of the quality of its fits across all three collision systems. A similar approach was presented in \cite{pp_SCE_ALICE, Vytautas_Kalweit}, with a notable difference in the manner in which the pion yields were calculated in the thermal model. In our case, the pion, and all other, yields are based on experimental values measured over a single unit of rapidity around mid-rapidity. Additionally, Ref. \cite{Cleymans_2021} also presented an independent SCE HRG model analysis with and without the use of S-matrix corrections. In the context of the present letter, this latter approach is not employed. 

For our SCE analysis, the $\pi K p \Lambda \Xi \Omega K_{\rm{S}}^{0} \phi$ yields are calculated by fixing the temperature to the flavour specific freeze-out temperatures at $\mu_{\rm{B}} = 0$ from our original study \cite{Flor_etal_PLB20}, such that $T_\mathrm{L} = 150$ MeV for $\pi  p$ and $T_\mathrm{S} = 165$ MeV for $ K \Lambda \Xi \Omega K_{\rm{S}}^{0} \phi$. We calculate the anti-hadron to $\pi^{+}$ ratio as a function of $\big \langle dN_\mathrm{ch}/d\eta \big \rangle$ and compare our results with the experimental data. The choice of the anti-hadron to $\pi^{+}$ ratio is made, as in \cite{RB_SequentialFO}, in order to use only particles produced during the evolution of the fireball. This allows us to facilitate future multiplicity dependent comparisons to lower collision energy measurements, where $\mu_\mathrm{B} \neq 0$. In order to explicitly show the temperature dependence of the fireball volumes, we calculate the volumes across all three systems as a function of $\big \langle dN_\mathrm{ch}/d\eta \big \rangle$ with temperatures fixed to the aforementioned flavour specific temperatures, as well as $T = 158$ MeV for a non-flavour-dependent temperature.

\section{Results and Discussion}
\begin{figure}[htbp]
\centering
\includegraphics[width=1.00\linewidth, trim = {0 0 0 0}, clip]{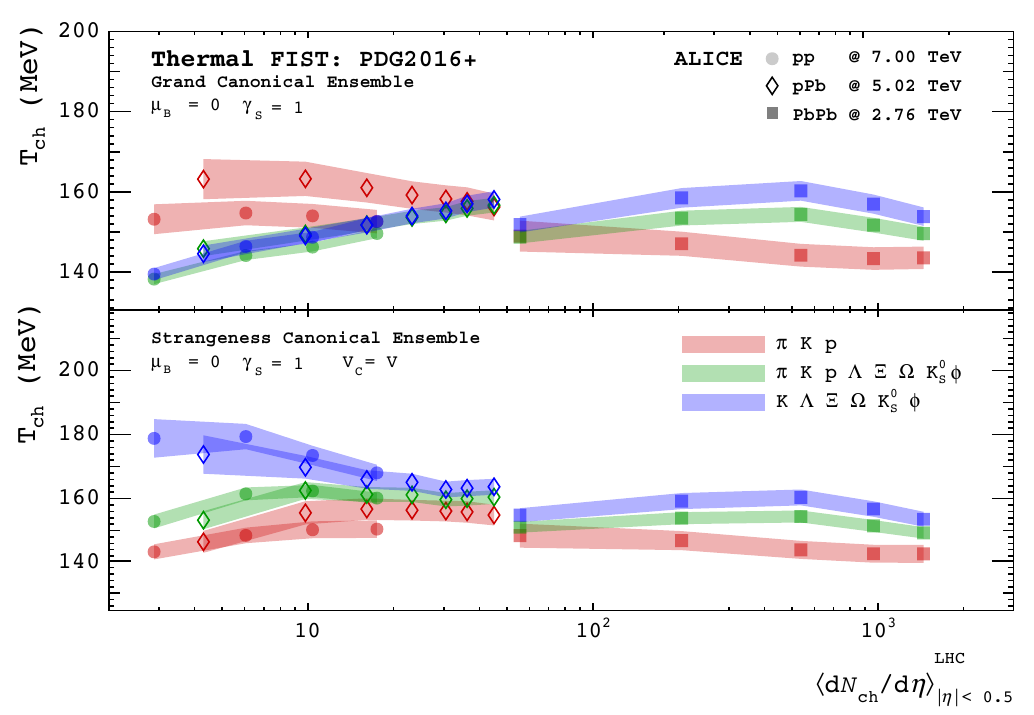}
\caption{Top Panel: Flavour-dependent Thermal-FIST GCE Fits to Yields using the PDG2016\Plus{} hadronic spectrum for ALICE pp collisions at $\sqrt{s} = 7.00$ TeV, pPb collisions at $\sqrt{s_{\rm{NN}}} = 5.02$ TeV and PbPb collisions at $\sqrt{s_{\rm{NN}}} = 2.76$ TeV, respectively, shown as as closed circles, open diamonds and closed squares as a function of $\big \langle dN_\mathrm{ch}/d\eta \big \rangle$. For all fits, $\mu_\mathrm{B} = 0$ and $\gamma_\mathrm{S} = 1$, while  $T_\mathrm{ch}$, and $V$ are used as free parameters. Red, green, and blue points represent the light, all, and strange fits, respectively. Bottom Panel: Flavour-dependent Thermal-FIST SCE Fits to Yields using the PDG2016\Plus{} hadronic spectrum for ALICE pp collisions at $\sqrt{s} = 7.00$ TeV, pPb collisions at $\sqrt{s_{\rm{NN}}} = 5.02$ TeV and PbPb collisions at $\sqrt{s_{\rm{NN}}} = 2.76$ TeV as a function of $\big \langle dN_\mathrm{ch}/d\eta \big \rangle$. For all SCE fits, $\mu_\mathrm{B} = 0$, $\gamma_\mathrm{S} = 1$ and $V_\mathrm{C} = V$, while  $T_\mathrm{ch}$, and $V$ are used as free parameters. The bottom panel follows the same labeling convention used in the top panel.}
\label{fig:GCE_SCE_T_Fits}
\end{figure}

\begin{figure}[htbp]
\centering
\includegraphics[width=1.0\linewidth, trim = {0 0 0 0}, clip]{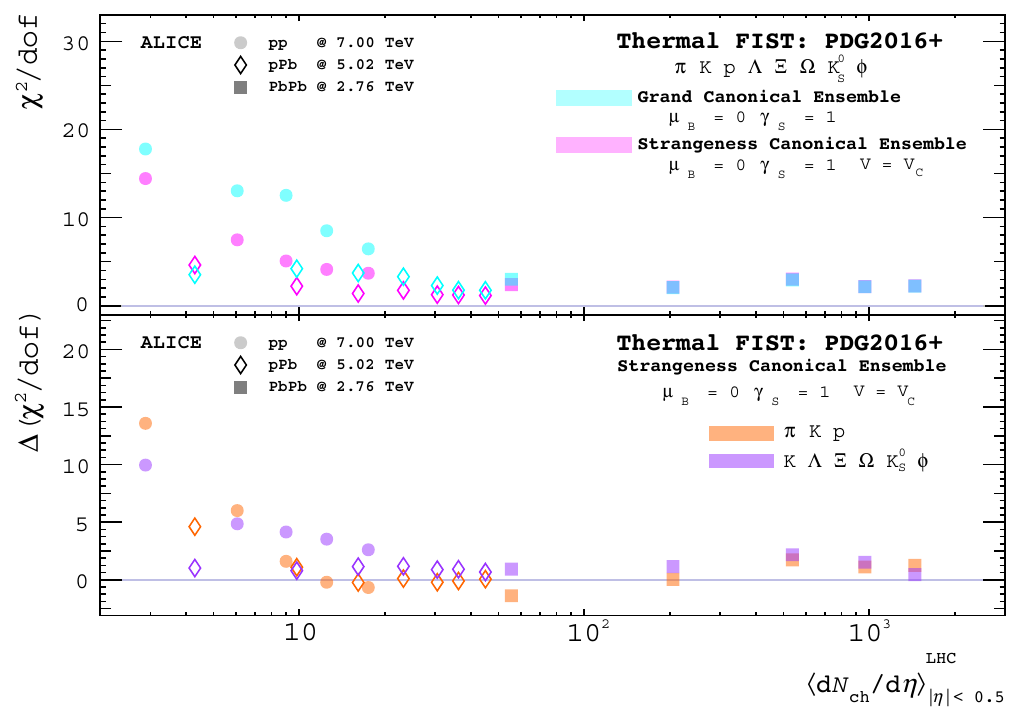}
\caption{Top Panel: Thermal-FIST $\chi^{2}$/dof values of the GCE (cyan) and SCE (magenta) fits used in the all fits of the top and bottom panels of Figure \cref{fig:GCE_SCE_T_Fits}, respectively, using the PDG2016\Plus{} hadronic spectrum for ALICE pp collisions at $\sqrt{s} = 7.00$ TeV, pPb collisions at $\sqrt{s_{\rm{NN}}} = 5.02$ TeV and PbPb collisions at $\sqrt{s_{\rm{NN}}} = 2.76$ TeV as a function of $\big \langle dN_\mathrm{ch}/d\eta \big \rangle$. Bottom Panel: Thermal-FIST $\Delta$($\chi^{2}$/dof) values for the SCE fits used in the light (orange) and strange (purple) fits of the bottom panel of Figure \cref{fig:GCE_SCE_T_Fits} via the PDG2016\Plus{} hadronic spectrum for ALICE pp collisions at $\sqrt{s} = 7.00$ TeV, pPb collisions at $\sqrt{s_{\rm{NN}}} = 5.02$ TeV and PbPb collisions at $\sqrt{s_{\rm{NN}}} = 2.76$ TeV as a function of $\big \langle dN_\mathrm{ch}/d\eta \big \rangle$. For both the top and bottom panels, the corresponding $\chi^{2}$/dof and $\Delta$($\chi^{2}$/dof) values for pp, pPb and PbPb collision systems are respectively shown as as closed circles, open diamonds and closed squares.}
\label{fig:ChiSq_DeltaChiSq}
\end{figure}

We extract freeze-out temperatures $T_{\mathrm{ch}}$ via Thermal-FIST for the light, all, and strange  particle thermal fits across increasing $\big \langle dN_\mathrm{ch}/d\eta \big \rangle$ values for pp, pPb, and PbPb collision systems at ALICE in the GCE and SCE configurations using experimental particle yields.

\begin{table*}[!h]
\caption{Thermal-FIST Grand Canonical Ensemble Yield Fits via the PDG2016\Plus{} hadronic spectrum for the most central (0 - 10\%) PbPb collisions at $\sqrt{s_{\rm{NN}}} = 2.76$ TeV. The top panel lists our previous results from Ref. \cite{Flor_etal_PLB20} whilst the bottom panel shows our results for the alternative light ($\pi p d$), full ($\pi p \Lambda \Xi \Omega K^0_S \phi d$), and strange ($K \Lambda \Xi \Omega K^0_S \phi$) particle fits, respectively, using the (anti-)deuteron yields from Ref. \cite{ALICEDeuterons}. For all fits, $\mu_B = 0$, whilst $T_\mathrm{ch}$ and $V$ are used as free parameters.}
\begin{center}
\setlength{\tabcolsep}{19pt}
\renewcommand{\arraystretch}{1}
\begin{tabular}{@{}cccc@{}}
\toprule
 Fit  & $T_{\mathrm{ch}}$ (MeV)  & $V(fm^3)$  & $\chi^{2}/dof$ \\
\midrule
 $\pi K p$ &  143.2 $\pm$ 2.79  & 8031.7 $\pm$ 1263 & 5.65/4 \\ 
 $\pi K p \Lambda \Xi \Omega K^0_S \phi$ &  149.6 $\pm$ 1.76  & 5764.4 $\pm$ 635.8 & 23.4/12 \\ 
$K \Lambda \Xi \Omega K^0_S \phi$ &  153.9 $\pm$ 2.30  & 4389.7 $\pm$ 640.8 & 10.5/8 \\ 

\midrule
\midrule

 Fit &  $T_\mathrm{ch}$ (MeV)  & $V(fm^3)$  & $\chi^{2}/dof$ \\
 \midrule
 $\pi p d$ & 144.6 $\pm$ 2.39  & 7911.6 $\pm$ 1177 & 5.45/4 \\ 
 $\pi K p \Lambda \Xi \Omega K^0_S \phi d$ & 150.1 $\pm$ 1.65  & 5613.6 $\pm$ 588.5 & 23.9/14 \\ 
 $K \Lambda \Xi \Omega K^0_S \phi$ & 153.9 $\pm$ 2.30  & 4389.7 $\pm$ 640.8 & 10.5/8 \\ % NO CHANGE
 \bottomrule
\end{tabular}
\end{center}
\label{tab:thermfitdeu}
\end{table*}

The top panel of Figure \cref{fig:GCE_SCE_T_Fits} shows the extracted freeze-out temperatures $T_{\mathrm{ch}}$ as a function of  $\big \langle dN_\mathrm{ch}/d\eta \big \rangle$ at $|\eta|<0.5$ for the three different fits within the GCE treatment. The pp, pPb, PbPb points are shown as closed circles, open diamonds and closed squares, respectively. The red, green, and blue points pertain to the extracted $T_{\mathrm{ch}}$ values for the light, all, and strange fits, respectively. The coloured bands represent the retained uncertainties from the fits onto the freeze-out parameters. The bottom panel of Figure \cref{fig:GCE_SCE_T_Fits} shows the extracted freeze-out temperatures $T_{\mathrm{ch}}$ as a function of  $\big \langle dN_\mathrm{ch}/d\eta \big \rangle$ at $|\eta|<0.5$ for the three different fits within the SCE treatment, following the same labeling convention as in the top panel of the figure. Qualitatively contrasting both top and bottom panels, we observe a deterioration of the flavour-dependent 2CFO approach occurring in the pp and pPb systems when using the GCE scenario, starting at values of $\big \langle dN_\mathrm{ch}/d\eta \big \rangle < 50$. This is not the case in the SCE treatment, where a flavour-dependent temperature separation is consistently present as a function of $\big \langle dN_\mathrm{ch}/d\eta \big \rangle$ across all three systems. It also is worth noting that in the grand canonical limit, i.e. for values of $\big \langle dN_\mathrm{ch}/d\eta \big \rangle > 50$, both the GCE and SCE treatments render almost identical $T_{\mathrm{ch}}$ values. Nevertheless, in order to qualitatively differentiate between the two treatments, we perform an in-depth comparison of the fit quality of the aforementioned $T_{\mathrm{ch}}$ values as a function of $\big \langle dN_\mathrm{ch}/d\eta \big \rangle$, which is separately shown in Figure \cref{fig:ChiSq_DeltaChiSq}. 

The top panel of Figure \cref{fig:ChiSq_DeltaChiSq} depicts the corresponding $\chi^{2}$/dof values for the GCE and SCE all fits of the top and bottom panels of Figure \cref{fig:GCE_SCE_T_Fits}, shown in cyan and magenta, respectively. The pp, pPb, PbPb points are shown as closed circles, open diamonds and closed squares. In the large system (PbPb), we observe a consistent quality of fit for all multiplicity classes in the GCE and SCE scenarios, with almost identical values for both approaches. However, in the GCE case, we observe a deteriorating quality of fits in the small systems (pp and pPb), with $\chi^2/\rm{dof}>5$ for nearly all fits to pp data. This renders an inconsistent description of temperatures across multiplicity within the GCE treatment of smaller systems. Instead, using the SCE approach, we observe a consistent quality of fit, specifically, $\chi^2/$dof$\;\lesssim 5$, across all small and large systems with corresponding values $\big \langle dN_{\rm{ch}}/d \eta\big\rangle \geq 10$.

The bottom panel of Figure \cref{fig:ChiSq_DeltaChiSq} shows the $\Delta$($\chi^{2}$/dof) values corresponding to the SCE fits performed in the light and strange fits from the bottom panel of Figure \cref{fig:GCE_SCE_T_Fits} -- shown in orange and purple, respectively. As in the top panel, the pp, pPb, PbPb points are respectively shown as closed circles, open diamonds and closed squares. The $\Delta$($\chi^{2}$/dof) values for the orange points in the bottom panel of Figure \cref{fig:ChiSq_DeltaChiSq} are calculated as the difference between the corresponding $\chi^{2}$/dof values of the SCE all (green) fits and the SCE light (red) fits from the bottom panel in Figure \cref{fig:GCE_SCE_T_Fits}.  Similarly, the $\Delta$($\chi^{2}$/dof) values for the purple points in the bottom panel of Figure \cref{fig:ChiSq_DeltaChiSq} are calculated as the difference between the corresponding $\chi^{2}$/dof values of the SCE all (green) fits and the SCE strange (blue) fits from the bottom panel in Figure \cref{fig:GCE_SCE_T_Fits}. This procedure is done in order to explicitly discriminate between notable differences in the quality of the fits when employing the 2CFO approach within the SCE, where any improvements of the fit quality would render $\Delta$($\chi^{2}$/dof) values approximately equal to the magenta $\chi^{2}$/dof values shown in the top panel. Conversely, any $\Delta$($\chi^{2}$/dof) values below zero reflect a worsened fit quality. We note that for all three systems, across all values of $\big \langle dN_{\rm{ch}}/d \eta\big\rangle$, the calculated $\Delta$($\chi^{2}$/dof) values for both the light (orange) and strange (purple) points generally follow the same trend as the all (cyan) points in the top panel. Without loss of generality, this suggests an overall improvement of the fits when employing the flavour-dependent 2CFO approach in the SCE treatment. The presence of negative $\Delta$($\chi^{2}$/dof) values can only be seen for a few of the light fits and can be attributed to a decreasing number of degrees of freedom, which causes an overall increase in the calculated $\chi^{2}$/dof. Moreover, the final $\chi^{2}$/dof values for the flavour specific fits can be obtained by subtracting the values in the lower panel from the magenta points in the upper panel of Figure \cref{fig:ChiSq_DeltaChiSq}. One can thus observe a consistent quality of fit -- specifically, $\chi^2/$dof$\;\lesssim 5$ -- for each of the three types of fits across the small and large systems with corresponding values of $\big \langle dN_{\rm{ch}}/d \eta\big\rangle \geq 10$. We note that the quality of fit is consistent for both the light and strange fits to a lower corresponding value of $\big \langle dN_{\rm{ch}}/d \eta\big\rangle \geq 5$. Generally, the $\chi^2/$dof values for the light and strange fits are lower than those for the all fit.  

\begin{figure}[htbp]
\centering
% trim is {left bottom right top}
\includegraphics[width=1.00\linewidth, trim = {0 0 0 0}, clip]{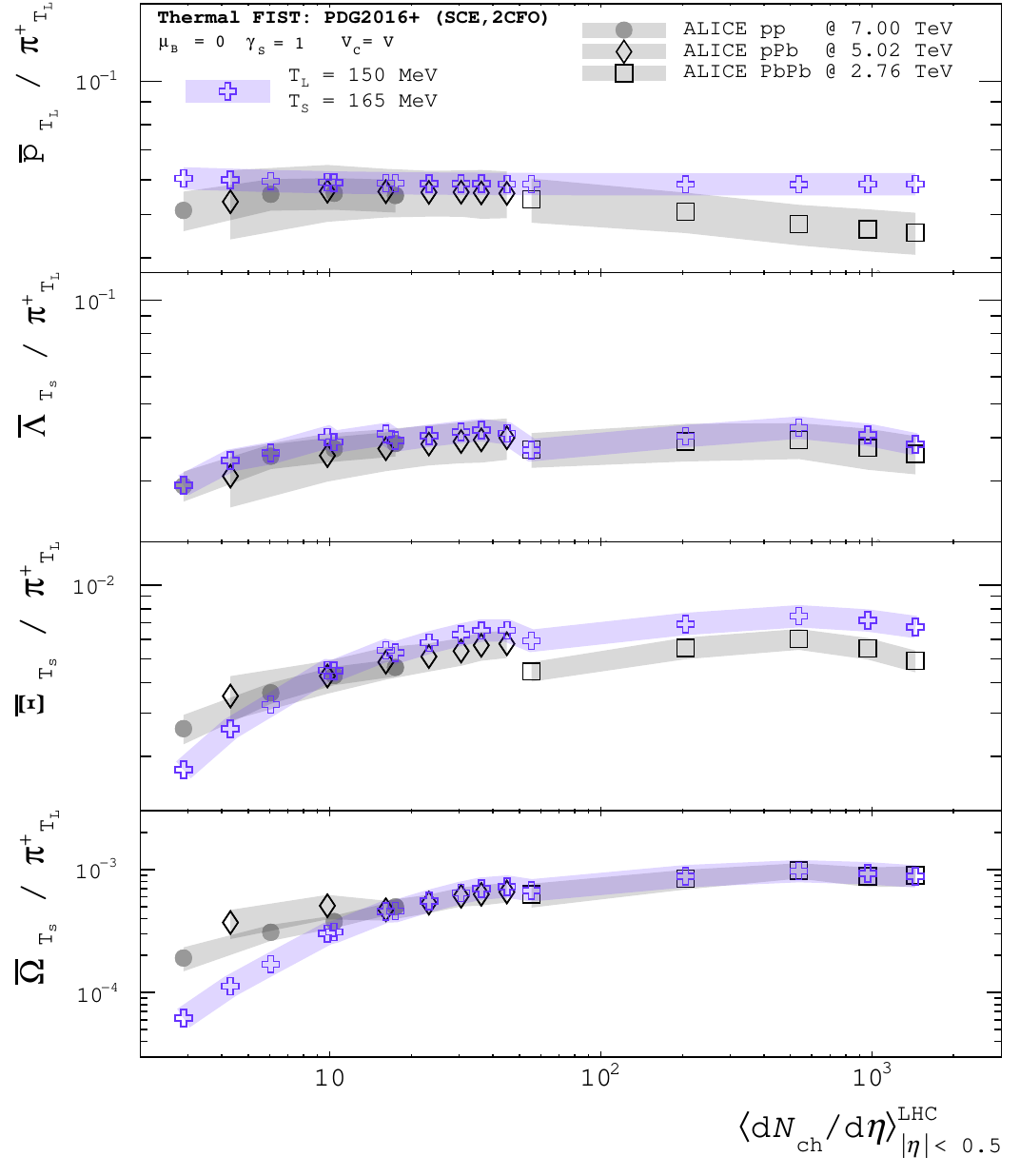}
\caption{2CFO SCE Thermal-FIST Thermal Model anti-hadron to $\pi^{+}$ yield ratio calculations via the PDG2016\Plus{} hadronic spectrum for ALICE pp collisions at $\sqrt{s} = 7.00$ TeV, pPb collisions at $\sqrt{s_{\rm{NN}}} = 5.02$ TeV, and PbPb collisions at $\sqrt{s_{\rm{NN}}} = 2.76$ TeV as a function of $\big \langle dN_\mathrm{ch}/d\eta\big \rangle$. From top to bottom: $\bar{p}$/$\pi^{+}$, $\bar{\Lambda}$/$\pi^{+}$, $\bar{\Xi}$/$\pi^{+}$, and $\bar{\Omega}$/$\pi^{+}$, respectively. ALICE experimental points for pp, pPb and both PbPb collision systems are respectively shown as grey closed circles, black open diamonds and black closed squares. Purple open crosses depict to our calculated SCE 2CFO anti-hadron to $\pi^{+}$ ratios at vanishing baryo-chemical potential in the SCE framework based on the flavour specific temperatures extracted in \cite{Flor_etal_PLB20}. For all calculations, $\mu_\mathrm{B} = 0$, $\gamma_\mathrm{S} = 1$ and $V_\mathrm{C} = V$.}
\label{fig:Yields_SCE_Fits_2CFO}
\end{figure}

\begin{figure}[htbp]
\centering
% trim is {left bottom right top}
\includegraphics[width=1.00\linewidth, trim = {0 0 0 0}, clip]{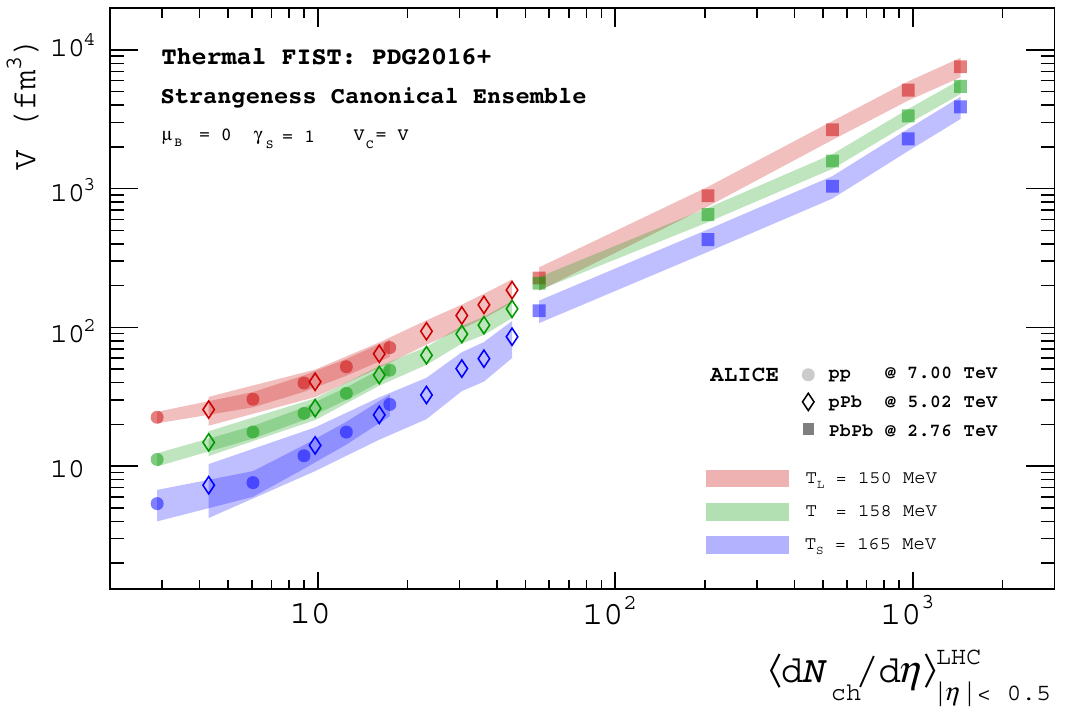}
\caption{Thermal-FIST SCE Thermal Model temperature dependent volume per unit rapidity calculations via the PDG2016\Plus{} hadronic spectrum for ALICE pp collisions at $\sqrt{s} = 7.00$ TeV, pPb collisions at $\sqrt{s_{\rm{NN}}} = 5.02$ TeV, and PbPb collisions at $\sqrt{s_{\rm{NN}}} = 2.76$ TeV, respectively shown as as closed circles, open diamonds and closed squares as a function of $\big \langle dN_\mathrm{ch}/d\eta \big \rangle$. The red, green, and blue points represent our volume calculations based on the flavour specific temperatures extracted in \cite{Flor_etal_PLB20} at 150 MeV, 158 MeV and 165 MeV, respectively. For all calculations, $\mu_\mathrm{B} = 0$, $\gamma_\mathrm{S} = 1$ and $V_\mathrm{C} = V$.}
\label{fig:Volumes_SCE}
\end{figure}

\begin{figure}[htbp]
\centering
% trim is {left bottom right top}
\includegraphics[width=1.0\linewidth, trim = {0 0 0 0}, clip]{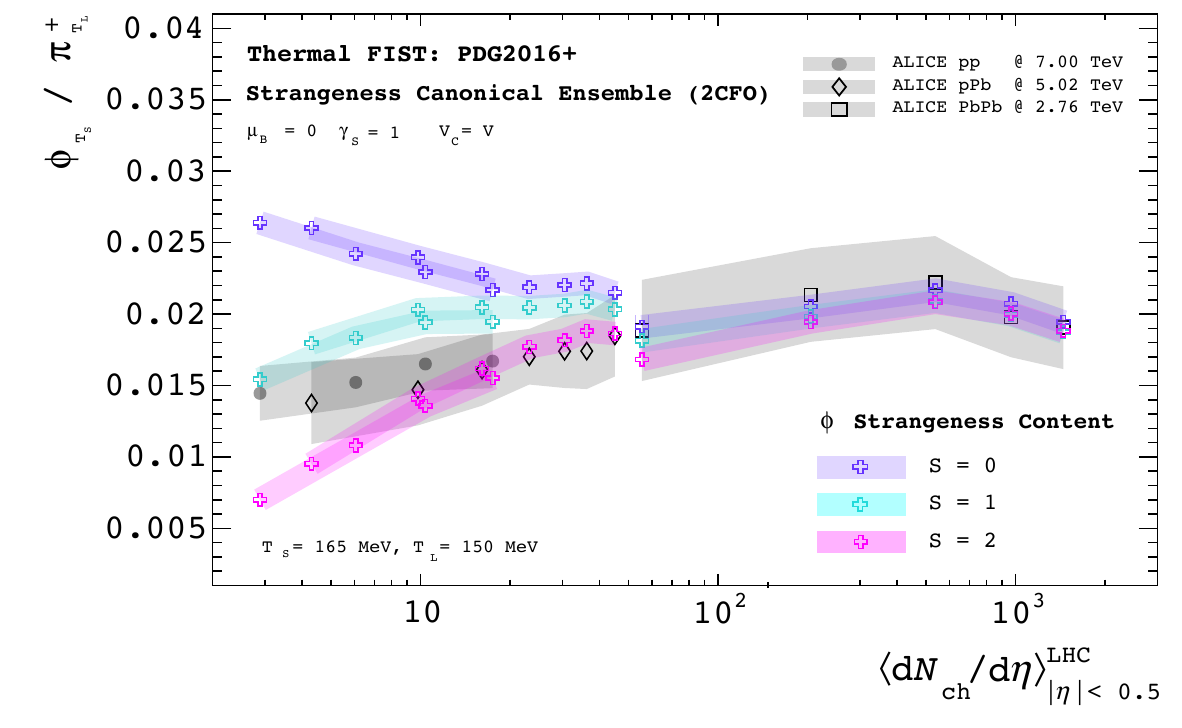}

\caption{2CFO SCE Thermal-FIST Thermal Model $\phi$ to $\pi^{+}$ yield ratio calculations via the PDG2016\Plus{} hadronic spectrum for ALICE pp collisions at $\sqrt{s} = 7.00$ TeV, pPb collisions at $\sqrt{s_{\rm{NN}}} = 5.02$ TeV, and PbPb collisions at $\sqrt{s_{\rm{NN}}} = 2.76$ TeV as a function of $\langle dN_\mathrm{ch}/d\eta\rangle$. ALICE experimental points for pp, pPb, PbPb collisions are respectively shown in grey as closed circles, open diamonds, and closed squares. Purple, cyan and magenta open crosses depict our calculated SCE 2CFO $\phi$ to $\pi^{+}$ ratios in the SCE framework based on the flavour specific freeze-out temperatures at vanishing baryo-chemical potential shown in \cite{Flor_etal_PLB20} fixing the total strangeness values for the $\phi$ meson to $S=0$, $S=1$ and $S=2$, respectively. For all calculations, $\mu_\mathrm{B} = 0$, $\gamma_\mathrm{S} = 1$ and $V_\mathrm{C} = V$.}
\label{fig:phi_fits}
\end{figure}

Figure \cref{fig:Yields_SCE_Fits_2CFO} shows the anti-hadron to $\pi^{+}$ ratio as a function of $\big \langle dN_\mathrm{ch}/d\eta\big \rangle$, starting from the top panel down to the bottom: $\bar{p}$/$\pi^{+}$, $\bar{\Lambda}$/$\pi^{+}$, $\bar{\Xi}$/$\pi^{+}$ and $\bar{\Omega}$/$\pi^{+}$, respectively. The experimental points for ALICE pp, pPb and both PbPb collision systems are shown as grey closed circles, black open diamonds and black closed squares, respectively. The purple open crosses indicate our calculated 2CFO anti-hadron to $\pi^{+}$ ratios within the SCE framework in Thermal-FIST using the flavour specific temperatures $T_\mathrm{L} = 150$ MeV for $\pi^{+}$ and $\bar{p}$, and $T_\mathrm{S} = 165$ MeV for $\bar{\Lambda} \bar{\Xi}$ and $\bar{\Omega}$. The corresponding model calculations of the $\phi$ to $\pi^{+}$ ratio are shown separately in Figure \cref{fig:phi_fits}, keeping the same purple cross colour convention, for the traditional treatment of net-strangeness of the $\phi$ meson of $S=0$. For this study we choose the extrapolated chemical freeze-out temperatures determined by the spline fits in our previous letter \cite{Flor_etal_PLB20} for central (0 - 10\%) PbPb collisions at $\mu_\mathrm{B}$ = 0 uniformly across all centralities and system sizes, namely, $T_\mathrm{L} = 150$ MeV and $T_\mathrm{S} = 165$ MeV. It is worth mentioning that an alternative use of $T_\mathrm{L} = 142$ MeV rather than $150$ MeV does not change the qualitative results shown in Figure \cref{fig:Yields_SCE_Fits_2CFO} -- particularly the $\bar{p}$/$\pi^{+}$ ratio presented in the top panel. We also test that using instead the temperatures shown in the bottom panel of Figure \cref{fig:GCE_SCE_T_Fits} does not make a difference in final state particle yield ratios since a simultaneous change in volume from the values shown in Figure \cref{fig:Volumes_SCE} will compensate any temperature differences.

Our results show an excellent agreement with the experimental yield ratios measured by ALICE across all three systems and are consistent with those shown in Refs. \cite{pp_SCE_ALICE, Vytautas_Kalweit, Cleymans_2021}. At these high collision energies, strangeness seems saturated even in the smallest systems. The main differences to Ref. \cite{Vytautas_Kalweit} emerge from the fact that our results do not require different rapidity windows for pions and strange particles and no additional normalization factor is used to reproduce the experimental results. In our analysis, the accurate representation of final state particle yields relies solely on the use of flavour-dependent temperatures and fireball volumes -- the relevant volumes as a function of $\big \langle dN_\mathrm{ch}/d\eta\big \rangle$ as determined by the model are shown in Figure \cref{fig:Volumes_SCE}.

Lastly, we vary the treatment of net-strangeness of the $\phi$ meson in our model between $S=0$,  $S=1$ and $S=2$ to gauge the sensitivity of the calculated yields and to elucidate the question of strangeness enhancement in the case of the vector meson. 
Figure \cref{fig:phi_fits} shows the $\phi$ to $\pi^{+}$ ratio as a function of  $\big \langle dN_\mathrm{ch}/d\eta \big \rangle$, following the same labeling convention for experimental points and 2CFO thermal model calculations as Figure \cref{fig:Yields_SCE_Fits_2CFO}, using a variation of the total strangeness (S) value within the model for the $\phi$ meson. To reiterate, the $\pi^{+}$ and $\phi$ values are calculated at $T_\mathrm{L} = 150$ MeV and $T_\mathrm{S} = 165$ MeV, respectively. For the comparison to the data, we arbitrarily fix the total strangeness of the $\phi$ meson to $S = 0$ (purple open crosses), $S = 1$ (cyan open crosses) and $S = 2$ (magenta open crosses), respectively. 

In the case of $S = 0$, the calculated yield ratios are well described by the 2CFO approach only in the large systems, for $\langle dN_\mathrm{ch}/d\eta\rangle$ > 50. Our model vastly overestimates the value of the ratio for both the pp and pPb systems. In the case  $S = 1$, we see an improvement particularly in the smaller systems, with our calculated values almost falling within the errors of the experimental data for all multiplicity bins. Lastly, for the case of $S = 2$, the model underestimates the experimental yield ratios for values of $\langle dN_\mathrm{ch}/d\eta\rangle$ < 10, but does otherwise quite well. Our results suggest that for $\phi$ production in small systems the $\phi$ should not be considered a $S = 0$ particle, because simple flavour conservation and recombination arguments require more than a single string to fragment to form an $s\bar{s}$ state. Therefore, $\phi$ yields can be more accurately calculated, within the HRG framework, by assuming a non-zero strangeness content for the $\phi$-meson.

\section{Conclusion}
We present determinations of freeze-out temperatures $T_{\mathrm{ch}}$ for the light, full, and strange particle thermal fits across increasing  $ \big \langle dN_\mathrm{ch}/d\eta \big \rangle$ values for pp, pPb, and PbPb collision systems at ALICE in the GCE and SCE configurations from experimental particle yields via Thermal-FIST using the PDG2016\Plus{} hadronic spectrum. Moreover, we also show thermal model anti-hadron to $\pi^{+}$ yield ratio calculations in these same collision systems as a function $\big \langle dN_\mathrm{ch}/d\eta \big \rangle$, with particular attention given to the treatment of the total strangeness content of the $\phi$ meson. In the scope of the Strangeness Canonical Ensemble within the framework of the Thermal-FIST HRG model package, we show an excellent description of experimental yield ratios across all three systems, measured by the ALICE Collaboration as a function of $\big \langle dN_\mathrm{ch}/d\eta \big \rangle$ at LHC energies, when employing flavour-dependent chemical freeze-out temperatures under the assumption of fully saturated strangeness. On the other hand, it should be noted that Grand Canonical Ensemble calculations with a sizeable $\gamma_{S}$ factor \cite{Chatterjee_2017}, Canonical Ensemble calculations with large correlation volumes \cite{VV_Donigus_Stoeker_PRC1}, and a dynamical core-corona initialization framework with large non-equilibrium contributions \cite{CoreCorona} also describe the experimental data, however, our approach is the only one showing a common particle production and flavour-dependent chemical freeze-out scenario consistently from the smallest to the largest collision systems at LHC energies. 

In conclusion, the flavour-dependent (T$_{ch}$) separation established in heavy-ion collisions, seems to prevail also at low $\langle dN_\mathrm{ch}/d\eta\rangle$ values corresponding to the pp and pPb systems. This sustained separation may also be seen as an indication of QGP formation in small systems.

\section{Acknowledgments}
The authors acknowledge fruitful discussions with Volodymyr Vovchenko, Claudia Ratti, Paolo Parotto,  Jamie Stafford, Livio Bianchi, Boris Hippolyte, Horst Sebastian Scheid, Dhevan Gangadharan and Helen Caines. This work was supported by the DOE grants DEFG02-07ER4152 and DE-SC004168. F.A.F. acknowledges ongoing support from the National Science Foundation Grant No. 2138010 and previous support from the Franco-American Fulbright Commission.

\bibliographystyle{apsrev4-1}
\bibliography{refs}

\end{document}